\documentclass[twocolumn,twoside,slac_two]{revtex4}
\def\jpsi{J/\psi}
\def\psip{\psi'}
\def\ch1{\chi_{c1}}
\def\c2{\chi_{c2}}
\def\1p1{^1P_1}
\def\etac{\eta_c}
\def\etac2s{\eta_c(2S)}
\def\etacp{\eta_c'}

\def\to{\rightarrow}
\def\pbar{\overline{p}}
\def\pbarp{\overline{p}p}

\def\cbarc{\overline{c}c}
\def\ccbar{c\overline{c}}
\def\qqbar{q\overline{q}}
\def\DDBAR{D\overline{D}}
\def\ee{e^+e^-}
\usepackage{graphicx}
\usepackage{fancyhdr}
\begin{document}

\title{The PANDA Experiment at FAIR}

%

\author{Diego Bettoni}
\affiliation{Istituto Nazionale di Fisica Nucleare, 44100 Ferrara, Italy}

\begin{abstract}
The physics program of the future FAIR facility covers a wide
range of topics that address central issues of strong interactions and QCD.
The antiproton beam of unprecedented quality in the momentum range from
1 GeV/c to 15 GeV/c will allow the PANDA experiment
to make high precision, high statistics
measurements, which include
charmonium and open charm spectroscopy, the search for exotic hadrons
and the
study of in-medium modifications of hadron masses.
\end{abstract}

\maketitle

\thispagestyle{fancy}

\section{Introduction}

One of the most challenging and fascinating 
goals of modern physics is the achievement of a fully quantitative
understanding of the strong interaction, which is the subject of hadron physics.
Significant progress has been achieved over the past few years thanks to
considerable advances in experiment and theory. New experimental results
have stimulated a very intense theoretical activity and a refinement of
the theoretical tools. 

Still there are many fundamental questions which remain basically
unanswered.
Phenomena such as the confinement of quarks, the existence of glueballs 
and hybrids, the origin of the masses of hadrons in the context of the
breaking of chiral symmetry are long-standing puzzles and represent
the intellectual challenge in our attempt to understand the nature of the
strong interaction and of hadronic matter.

Experimentally, studies of hadron structure can be performed with different
probes such electrons, pions, kaons, protons or antiprotons.
In antiproton-proton annihilation particles with gluonic degrees of freedom
as well as particle-antiparticle pairs are copiously produced,
allowing spectroscopic studies with very high statistics and precision.
Therefore, antiprotons are an excellent tool to address the open problems.

The recently approved FAIR facility
(Facility for Antiproton and Ion Research), 
which will be built as a major upgrade
of the existing GSI laboratory in Germany, will provide antiproton beams
of the highest quality in terms of intensity and resolution, which will
provide an excellent tool to answer these fundamental questions.

The PANDA experiment (Pbar ANnihilations at DArmstadt) will use the
antiproton beam from the High-Energy Storage Ring (HESR) colliding with an
internal proton
target and a general purpose spectrometer to carry out a rich and
diversified hadron physics program, which includes
charmonium and open charm spectroscopy, the search for exotic hadrons
and the study of in-medium modifications of hadron masses.

This paper is organized as follows: in section 2 we will give an overview of
the FAIR facility and the HESR; in section 3 we will discuss some of the
most significant items of the PANDA experimental program; in section 4
we will give a brief description of the PANDA detector. 
Finally in section 5 we will present our conclusions.

\section{The FAIR facility}

\begin{figure}
\begin{center}
\includegraphics[width=80mm]{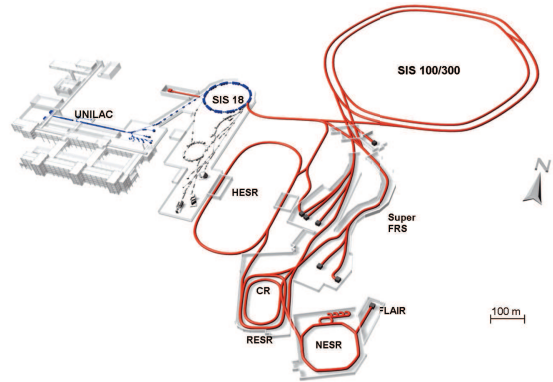}
\end{center}
\caption{The FAIR complex.}
\end{figure}

The planned FAIR complex is shown in Fig. 1.
The heart of the system consists of two synchrotron rings, called SIS100 and SIS300, 
housed in the same tunnel, which will provide proton and ion beams of 
unprecedented quality. 
The SIS100, a 100 t$\cdot$m proton ring, will feed the radioactive ion and
antiproton beam lines for experiments to be carried out in the High-Energy
Storage Ring (HESR), the Collector and Cooler rings (CR) and the New
Experimental Storage Ring (NESR). 
The SIS300 will deliver high energy ion beams for the study of
ultra relativistic heavy ion collisions.

The accelerators of FAIR will feature significant improvements in system
parameters over existing facilities:
\begin{itemize}
\item {\bf beam intensity}: increased by a factor of 100 to 1000 for primary
and 10000 for secondary beams;
\item {\bf beam energy} will increase by a factor 30 for heavy ions;
\item {\bf beam variety}: FAIR will offer a variety of beam lines, from
antiprotons to protons, to uranium and radioactive ions;
\item {\bf beam precision}: availability of cooled antiproton and ion beams
(stochastic and electron cooling);
\item {\bf parallel operation}: full accelerator performance for up to four
different, independent experiments and experimental programs.
\end{itemize} 

These features will make FAIR a first rate facility for experiments in
particle, nuclear, atomic, plasma and applied physics. 

\subsection{The High-Energy Storage Ring}

The antiproton beam will be produced by a primary proton beam from the SIS100.
The $\pbar$ production rate will be of approximately $2\times 10^7$/s.
After $5\times 10^5\ \pbar$ have been produced they will be transferred to the
HESR, where internal experiments in the $\pbar$ momentum range from 1 GeV/c to
15 GeV/c can be performed.

\begin{figure}
\label{fig:hesr}
\begin{center}
\includegraphics[width=80mm]{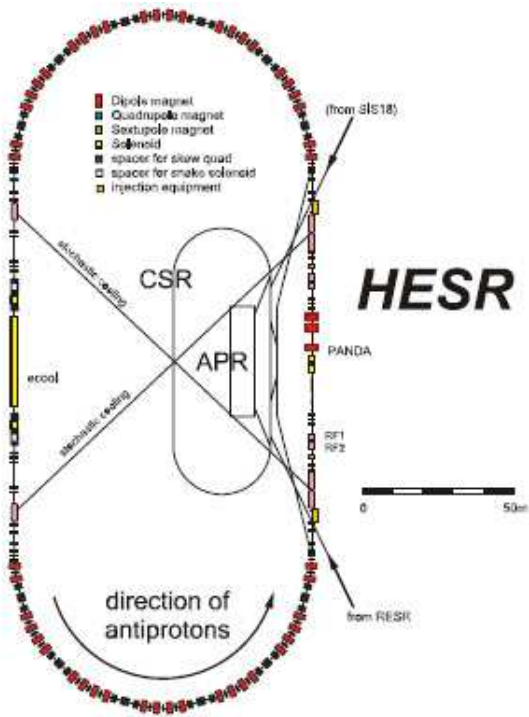}
\caption{Layout of the HESR.}
\end{center}
\end{figure}

The layout of the HESR is shown in Fig. 2.
It is a racetrack ring, 574 meters in length, with two straight sections which
will host the electron cooling and, respectively, the PANDA experiment.
Two modes of operation are foreseen: in the high-luminosity mode peak 
luminosities of $2 \times 10^{32}$cm$^{-2}$s$^{-1}$ will be reached with a beam
momentum spread $\delta p/p = 10^{-4}$, achieved by means of stochastic cooling;
in the high-resolution mode for beam momenta below 8 GeV/c electron cooling
will yield a smaller beam momentum spread $\delta p/p = 10^{-5}$ at a
reduced luminosity of $10^{31}$cm$^{-2}$s$^{-1}$.
The high-resolution mode will allow to measure directly the total width
of very narrow (below 1 MeV) resonances.

\section{The PANDA Physics Program}

The PANDA experiment has a rich experimental program whose ultimate aim is
to improve our knowledge of the strong interaction and of hadron structure.
The experiment is being designed to fully exploit the extraordinary 
physics potential arising from the availability of high-intensity, cooled
antiproton beams.
Significant progress beyond the present understanding of the field is expected
thanks to improvements in statistics and precision of the data.

Many experiments are foreseen in PANDA. In this paper we will discuss 
the following:
\begin{itemize}
\item charmonium spectroscopy;
\item search for gluonic excitations (hybrids and glueballs);
\item study of hadrons in nuclear matter;
\item open charm spectroscopy. 
\end{itemize}

\subsection{Charmonium Spectroscopy}
Ever since its discovery in 1974 \cite{psidisc} charmonium has been a powerful tool
for the understanding of the strong interaction.
The high mass of the $c$ quark (m$_c$ $\approx$ 1.5\ GeV/c$^2$)
makes it plausible to attempt a description of the dynamical properties
of the ($\ccbar$) system in terms of non-relativistic potential models,
in which the functional form of the potential is chosen to reproduce
the asymptotic properties of the strong interaction. The free
parameters in these models are to be determined from a comparison
with the experimental data.

Now, more than thirty years after the $\jpsi$ discovery, charmonium physics continues
to be an exciting and interesting field of research. 
The recent discoveries of new states ($\etacp$, X(3872)), 
and the exploitation of the B factories as rich sources of
charmonium states
have given rise to renewed interest in heavy quarkonia,
and stimulated a lot of experimental and theoretical activities.
Over the past few years a significant progress
has been achieved by Lattice Gauge Theory
calculations, which have become increasingly more capable of dealing
quantitatively with non perturbative dynamics in all its aspects,
starting from the first principles of QCD.

\subsubsection{Experimental Study of Charmonium}

Experimentally charmonium has been studied mainly in $\ee$ and $\pbarp$ experiments.

In $\ee$ annihilations direct charmonium formation is possible 
only for states with the quantum numbers of the 
photon $J^{PC}=1^{--}$, namely  the 
$\jpsi$, $\psip$ and $\psi(3770)$ resonances.
Precise measurements of the masses and widths of these states can be
obtained from the energy of the electron and positron beams, which
are known with good accuracy.
All other states can be reached by means of other production mechanisms,
such as photon-photon fusion, initial state radiation, B-meson decay
and double charmonium.

On the other hand all $\ccbar$ states can be directly formed in $\pbarp$
annihilations, through the coherent annihilation of the three quarks in the
proton with the three antiquarks in the antiproton. 
This technique, originally
proposed by P. Dalpiaz in 1979 \cite{dalpiaz79}, 
could be successfully employed a few years 
later at CERN and Fermilab thanks to the development of stochastic cooling.
With this method the masses and widths of all charmonium states can be measured 
with excellent accuracy, determined by the very precise knowledge of the initial
$\pbarp$ state and not limited by the resolution of the detector.

The cross section for the process:
\begin{equation}
\pbarp \to (\cbarc) \to {\rm final~state}
\end{equation}
is given (in units $\hbar=c=1$) by the well known Breit-Wigner formula:
\begin{equation}
\sigma_{BW}(E) = \frac{2J + 1}{4} \frac{\pi}{k^2}
\frac{B_{in} B_{out} \Gamma_{R}^{2}}{(E - M_R)^2 + 
\Gamma_{R}^{2}/4} \label{eq:bw}
\end{equation}
where $E$ and $k$ are the center-of-mass (c.m.) energy
and momentum; $J$, $M_R$ and $\Gamma_R$ are the
resonance spin, mass and total width and $B_{in}$ and
$B_{out}$ are the branching ratios into the initial
($\pbarp$) and final states.
Due to the finite energy spread of the beam, the measured
cross section is a convolution of the Breit-Wigner
cross section, eq.~(\ref{eq:bw}), and the beam energy
distribution function $f(E,\Delta E_B)$;
the effective production rate $\nu$ is given by:
\begin{equation}
\nu = L_0 \left\{ \epsilon \int dE f(E,\Delta E_B) 
\sigma_{BW}(E) + \sigma_b \right\}
\label{eq:prod}
\end{equation}
where $L_0$ is the instanteneous luminosity, $\epsilon$ an
overall efficiency $\times$ acceptance factor and
$\sigma_b$ a background term.

The parameters of a given resonance can be extracted by
measuring the formation rate for that resonance as a
function of the c.m. energy $E_{cm}$.
The accurate determination of masses and widths depends
crucially on the precise knowledge of the absolute energy
scale and on the beam energy spectrum.

\begin{figure}[htb]
\begin{center}
\includegraphics[width=80mm]{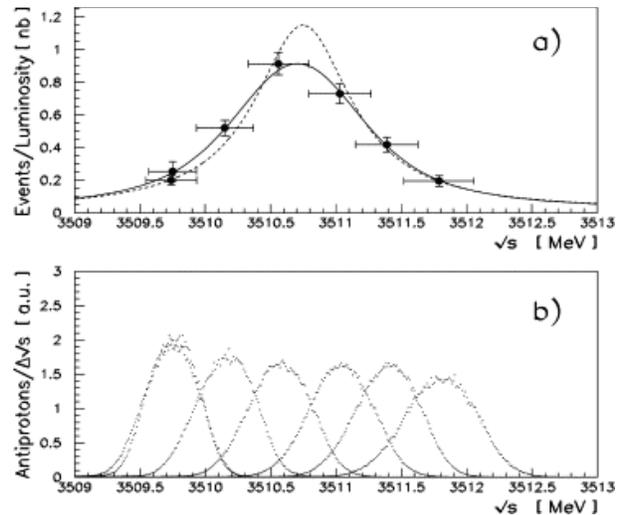}
\caption{Resonance scan at the $\ch1$ carried out at Fermilab (a) and beam energy
distribution in each data point (b). 
\label{fig:chi1scan}}
\end{center}
\end{figure}

The technique is illustrated in Fig.~\ref{fig:chi1scan} which shows a scan of the
$\ch1$ resonance carried out at the Fermilab antiproton accumulator by the E835
experiment~\cite{chi1E835} using the process $\pbarp \to \ch1 \to \jpsi\gamma$. 
For each point of the scan the horizontal error bar in (a)
corresponds to the width of the beam energy distribution. The actual beam energy
distribution is shown in (b). This scan allowed the E835 experiment to carry out the
most precise measurement of the mass ($3510.719 \pm 0.051 \pm 0.019$ Mev/c$^2$) and total
width ($0.876 \pm 0.045 \pm 0.026$ MeV) of this resonance.

\subsubsection{The charmonium spectrum}

\begin{figure}[htb]
\begin{center}
\includegraphics[width=80mm]{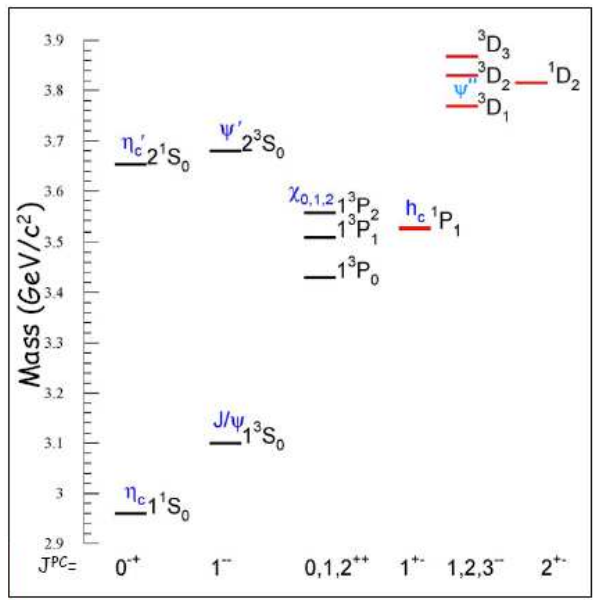}
\caption{The charmonium spectrum.
\label{fig:ccspectrum}}
\end{center}
\end{figure}

The spectrum of charmonium states is shown in Fig.~\ref{fig:ccspectrum}.
It consists of eight narrow states below the open charm threshold  (3.73 GeV) and 
several tens of states above the threshold.

All eight states {\bf below $\DDBAR$ threshold} are well established, but whereas 
the triplet states are measured with very good accuracy, the same cannot be said 
for the singlet states.

The $\eta_c$ was discovered almost thirty years ago and many measurements of its
mass and total width exist, with six new measurements in the last four years. Still
the situation is far from satisfactory. 
The Particle Data Group (PDG)~\cite{PDG} value of the mass 
is $2980.4 \pm 1.2$ MeV/c$^2$, an
average of eight measurements with an internal confidence level of 0.026: the
error on the $\eta_c$ mass is still as large as 1.2 MeV/c$^2$, to be compared with few
tens of KeV/c$^2$ for the $\jpsi$ and $\psip$ and few hundreds of KeV/c$^2$ for the $\chi_{cJ}$
states. The situation is even worse for the total width: the PDG average is 
$25.5 \pm 3.4$ MeV, with an overall confidence level of only 0.001 and individual
measurements ranging from 7 MeV to 34.3 MeV. The most recent measurements 
have shown that the $\eta_c$ width is larger than was previously believed, with
values which are difficult to accomodate in quark models. 
This situation points
to the need for new high-precision measurements of the $\eta_c$ parameters. 

The first experimental evidence of the $\etac2s$ was reported by the Crystal
Ball collaboration~\cite{etacpCB}, but this finding was not confirmed in
subsequent searches in $\pbarp$ or $\ee$ experiments. The $\etac2s$ was finally
discovered by the Belle collaboration~\cite{etacpBelle}
in the hadronic decay of the B meson
$B \to K + \etac2s \to K+ (K_sK^-\pi^+)$ with a mass which was incompatible with
the Crystal Ball candidate. The Belle finding was then confirmed by CLEO~\cite{etacpCLEO}
and BaBar~\cite{etacpbabar} which observed this state in two-photon fusion.
The PDG value of the mass is $3638 \pm 4$ MeV/c$^2$, corresponding to a surprisingly
small hyperfine splitting of $48 \pm 4$ MeV/c$^2$, whereas the total width is only 
measured with an accuracy of 50\%. The study of this state has just started and all
its properties need to be measured with good accuracy.

The $\1p1$ state of charmonium ($h_c$) is of particular importance in the
determination of the spin-dependent component of the $\qqbar$ confinement potential.
The Fermilab experiment E760 reported an $h_c$ candidate in the decay channel
$\jpsi\pi^0$~\cite{1p1E760}, with a mass of $3526.2 \pm 0.15 \pm 0.2$ MeV/c$^2$. 
This finding was not confirmed by the successor experiment
E835, which however observed an enhancement in the $\eta_c\gamma$~\cite{1p1E835}
final state at a mass
of $3525.8 \pm 0.2 \pm 0.2$ MeV/c$^2$. 
The $h_c$ was finally observed by the CLEO collaboration~\cite{1p1CLEO}
in the process
$\ee \to \psip \to h_c + \pi^0$ with $h_c \to \eta_c + \gamma$, in 
which the $\eta_c$ was identified via its hadronic decays. They
found a value for the mass of $3524.4 \pm 0.6 \pm 0.4$ MeV/c$^2$.
It is clear that the study of this state has just started and that many more
measurements will be needed to determine its properties, in particular the width.

The region {\bf above $\DDBAR$ threshold} is rich in interesting new physics. 
In this region, close to the $\DDBAR$ 
threshold, one expects to find the four 1D states. 
Of these only the $1^3D_1$, identified with the
$\psi$(3770) resonance, has been found.
The $J = 2$ 
states ($1^1D_2$ and $1^3D_2$) are predicted to be narrow, because
parity conservation forbids their decay to $\DDBAR$. 
In addition to the D states, the radial excitations of the S and
P states are predicted to occur above the open charm threshold.
None of these states have been positively identified.

The experimental knowledge of this energy region comes from data taken at the
early $\ee$ experiments at SLAC and DESY and, more recently, at the B-factories, CLEO-c
and BES.
The structures and the higher vector states observed by the early $\ee$ experiments
have not all been confirmed by the latest much more accurate 
measurements by BES~\cite{RBES}. 
A lot of new states have recently been discovered at the B-factories, mainly in
the hadronic decays of the meson: these new states (X, Y, Z ...) are associated
with charmonium because they decay predominantly into charmonium states such
as the $\jpsi$ or the $\psip$, but their interpretation is far from obvious.
The situation can be roughly summarized as follows:
\begin{itemize}
\item the Z(3931)~\cite{Z3931}, observed in two-photon fusion and decaying predominantly
 into $\DDBAR$, is tentatively identified with the $\c2$(2S);
\item the X(3940)~\cite{X3940}, observed in double charmonium events, is 
tentatively identified with the $\eta_c$(3S);
\item for all other new states (X(3872), Y(3940), Y(4260), Y(4320) and so on) the
interpretation is not at all clear, with speculations ranging from the missing
$\ccbar$ states, to molecules, tetraquark states, and hybrids. 
It is obvious that further measurements are needed to determine the nature
of these new resonances.
\end{itemize}

The main challenge of the next years will be thus to understand what these new 
states are and to match these experimental findings to the theoretical expectations
for charmonium above threshold. 

\subsubsection{Charmonium in PANDA}

Charmonium spectroscopy is one of the main items in the experimental program
of PANDA, and the design of the detector and of the accelerator are optimized
to be well suited for this kind of physics. 
PANDA will represent a substantial improvement over the Fermilab experiments
E760 and E835:
\begin{itemize}
\item up to ten times higher instantaneous luminosity 
(${\cal L} = 2 \times 10^{32}$cm$^{-2}$s$^{-1}$ in high-luminosity mode, compared
to $2 \times 10^{31}$cm$^{-2}$s$^{-1}$ at Fermilab);
\item better beam momentum resolution ($\Delta p/p = 10^{-5}$ in high-resolution
mode, compared with $10^{-4}$ at Fermilab);
\item a better detector (higher angular coverage, magnetic field, ability to
detect the hadronic decay modes).
\end{itemize}

At full luminosity PANDA will be able to collect several thousand $\ccbar$ states
per day.
By means of fine scans it will be possible to measure masses with accuracies of the
order of 100 KeV and widths to 10\% or better.
The entire energy region below and above open charm threshold will be explored.

\subsection{Gluonic Excitations}
One of the main challenges of hadron physics, and an important item in the PANDA
physics program, is the search for gluonic excitations, i.e. hadrons in which the
gluons can act as principal components.
These {\emph gluonic hadrons} fall into two main categories: glueballs, i.e.
states of pure glue, and hybrids, which consist of a $\qqbar$ pair and excited glue.
The additional degrees of freedom carried by gluons allow these hybrids and glueballs
to have $J^{PC}$ exotic quantum numbers: in this case mixing effects with nearby
$\qqbar$ states are excluded and this makes their experimental identification easier.
The properties of glueballs and hybrids are determined by the long-distance features
of QCD and their study will yield fundamental insight into the structure of the QCD
vacuum.

Antiproton-proton annihilations provide a very favourable environment in which
to look for gluonic hadrons. 
Two particles, first seen in $\pi N$ scattering~\cite{piN} with exotic quantum numbers
$J^{PC}=1^{-+}$, $\pi_1(1400)$~\cite{pi1400} and $\pi_1(1600)$~\cite{pi1600}, are
clearly seen in $\pbarp$ annihilation at rest.
On the other hand a narrow state at 1500 MeV/c$^2$ discovered in $\pbarp$ annihilations
by the Crystal Barrel experiment~\cite{f01500}, is considered the best candidate
for the glueball ground state ($J^{PC}=0^{++}$), even though the mixing with nearby
$\qqbar$ states makes this interpretation difficult.

So far the experimental search for glueballs and hybrids has been mainly 
carried out in the mass region below 2.2 MeV/c$^2$. 
PANDA will extend the search to higher masses and in particular to the 
charmonium mass region, were light quark states form a structure-less continuum
and heavy quark states are far fewer in number. 
Therefore exotic hadrons in this mass region could be resolved and identified
unambiguously.

\subsubsection{Charmonium Hybrids}

The spectrum of charmonium hybrid mesons can be calculated within the 
framework of various theoretical models, such as the bag model, the flux tube model,
the constituent quark model and recently, with increasing precision, from
Lattice QCD (LQCD). 
For these calculations the parameters are fixed according to the properties
of the known $\qqbar$ states.
All model predictions and LQCD calculations agree that the masses of the lowest
lying charmonium hybrids are between 4.2 GeV/c$^2$ and 4.5 GeV/c$^2$. Three of
these states are expected to have $J^{PC}$ exotic quantum numbers 
($0^{+-}$, $1^{-+}$, $2^{+-}$), making their experimental identification easier
since they will not mix with nearby $\ccbar$ states.
These states are expected to be narrower than conventional charmonium, because
their decay to open charm will be suppressed or forbidden below the 
$D\overline{D^*_J}$ threshold. 
The cross sections for the formation and production of charmonium hybrids are
estimated to be similar to those of normal charmonium states, which are within
experimental reach.
Formation experiments will generate only non-exotic charmonium hybrids, whereas
production experiments will yield both exotic and non-exotic states. 
This feature can be exploited experimentally: the observation of a state in 
production but not in formation will be, in itself, a strong hint of exotic behavior.

\subsubsection{Glueballs}

The glueball spectrum can be calculated within the framework of LQCD in the 
quenched approximation~\cite{morningstar}. 
In the mass range accessible to PANDA as many as 15 glueball states are
predicted, some with exotic quantum numbers (\emph{oddballs}).
As with hybrids, exotic glueballs are easier to identify experimentally
since they do not mix with conventional mesons. 
The complications arising from mixing with normal $\qqbar$ states is 
well illustrated by the case of the f$_0$(1500).
As mentioned above, this narrow state, observed at LEAR by the Crystal 
Barrel~\cite{f01500} and Obelix~\cite{f01500ob} experiments, is considered
the best candidate for the ground state glueball. 
However this interpretation is not unique, and relies on the combined analysis
of the complete set of two-body decays of the f$_0$(1500) and two other
scalar states, the f$_0$(1370) and the f$_0$(1710). This analysis yields 
the following mixing picture~\cite{f01500an}:
\begin{eqnarray}
|f_0(1710)> = 0.39|gg> + 0.91|s\overline{s}> + 0.14|N\overline{N}>  \\
|f_0(1500)> = -0.69|gg> + 0.37|s\overline{s}> - 0.62|N\overline{N}>\\
|f_0(1370)> = 0.60|gg> - 0.13|s\overline{s}> - 0.79|N\overline{N}>
\end{eqnarray}
where $|N\overline{N}> = (|u\overline{u}> + |d\overline{d}>)/\sqrt{2}$.
Other scenarios for the scalar meson nonet not involving a glueball have
been proposed and this makes the interpretation of the f$_0$(1500) as
the ground state glueball ambiguous.
This example highlights the need to extend the glueball search to higher
mass regions, which are free of the problem of mixing with conventional
$\qqbar$ states.

\subsection{Hadrons in Nuclear Matter}

The study of medium modifications of hadrons embedded in hadronic matter
is aimed at understanding the origin of hadron masses in the context
of spontaneous chiral symmetry breaking in QCD and its partial restoration
in a hadronic environment. 
So far experiments have been focussed on the light quark sector: evidence of
mass changes for pions and kaons have been deduced by the study of deeply
bound pionic atoms~\cite{pishift} and of K meson production in proton-nucleus
and heavy-ion collisions~\cite{kshift}.

The high-intensity $\pbar$ beam of up to 15\ GeV/c will allow an extension of
this program to the charm sector both for hadrons with hidden and open charm.
The in-medium masses of these states are expected to be affected primarily
by the gluon condensate.
Recent theoretical calculations predict small mass shifts (5-10 MeV/c$^2$)
for the low-lying charmonium states~\cite{TPR73} and more consistent effects
for the $\chi_{cJ}$ (40 MeV/c$^2$), $\psip$ (100 MeV/c$^2$) and 
$\psi$(3770) (140 MeV/c$^2$)~\cite{TPR7475}.  

D mesons, on the other hand, offer the unique opportunity to study the 
in-medium dynamics of a system with a single light quark. Recent theoretical
calculations agree in the prediction of a mass splitting for D mesons in 
nuclear matter but, unfortunately, they disagree in sign and size of the effect.

Experimentally the in-medium masses of charmonium states can be reconstructed
from their decay into di-leptons and photons, which are not affected by
final state interaction.
D meson masses, on the other hand, need to be reconstructed by their weak
decays into pions and kaons which makes the direct measurement of mass
modifications difficult. 
Therefore other signals have been proposed for the detection of in-medium
mass shifts of D mesons: in particular it has been speculated that
a lowering of the $\DDBAR$ threshold would result in an increased 
$D$ and $\overline{D}$ production in $\pbar$-nucleus annihilations~\cite{TPR78} 
or in an increase in width of the charmonium states 
lying close to the threshold~\cite{TPR7782}. 

Another study which can be carried out in PANDA is the measurement of $\jpsi$
and D meson production cross sections in $\pbar$ annihilation
on a series of nuclear targets. The
comparison of the resonant $\jpsi$ yield obtained from $\pbar$ annihilation
on protons and different nuclear targets allows to deduce the
$\jpsi$-nucleus dissociation cross section, a fundamental parameter to 
understand $\jpsi$ suppression in relativistic heavy ion collisions interpreted
as a signal for quark-gluon plasma formation. 

\section{Open Charm Physics}

The HESR running at full luminosity and at $\pbar$
momenta larger than 6.4\ GeV/c would produce a large number
of $D$ meson pairs.
The high yield (e.g. 100 charm pairs per second around the $\psi$(4040)) and
the well defined production kinematics of $D$ meson pairs would allow to
carry out a significant charmed meson spectroscopy program which would include,
for example, the rich $D$ and $D_s$ meson spectra.

\begin{figure}
\label{dspectrum}
\begin{center}
\includegraphics[width=80mm]{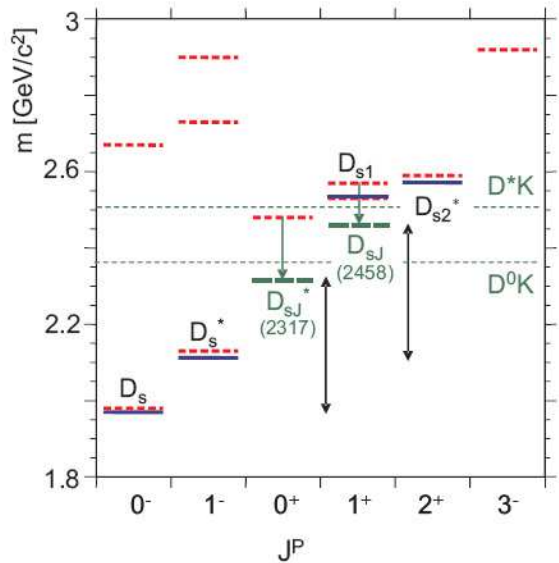}
\caption{The $D_s$ spectrum.}
\end{center}
\end{figure}

The B-factory experiments have discovered several new resonances in the
$D$ and $D_s$ sectors, where two are extremely narrow: the 
$D^\ast_{sJ}$(2317)~\cite{TPR95} and the $D^\ast_{sJ}$(2317)~\cite{TPR9698}.
These new states appear at unexpected locations, since their masses are more
than 140\ MeV/c$^2$ lower than expected from potential models,
as shown in Fig. 5.
This has given rise to speculations about their nature.
It is important to verify these findings by means of new measurements.
Threshold pair production can be employed for precision measurements of the mass
and the width of the narrow excited $D$ states. 

\section{The PANDA Detector}

In order to carry out the physics program discussed above the PANDA detector must
fulfil a number of requirements: it must provide (nearly) full solid angle coverage,
it must be able to handle high rates ($2 \times 10^7$ annihilations/s) with
good particle identification and momentum resolution for $\gamma$, e, $\mu$, $\pi$,
K and p. Additional requirements include vertex reconstruction capability and, for 
charmonium, a pointlike interaction region, efficient lepton identification and
excellent calorimetry (both in terms of resolution and of sensitivity to low-energy
showers).

\begin{figure*}[t]
\begin{center}
\includegraphics[width=135mm]{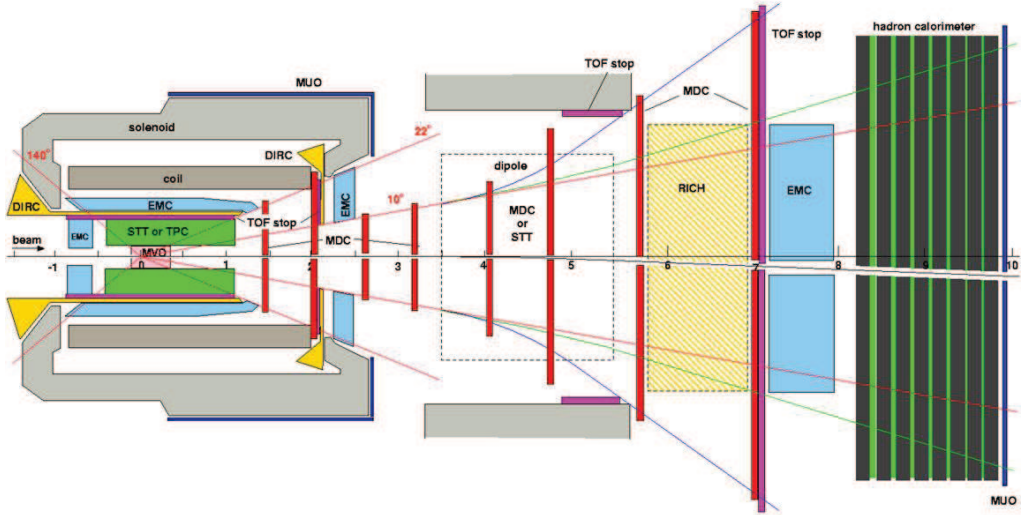}
\caption{Schematic view of the PANDA detector.} 
\label{fig:det}
\end{center}
\end{figure*}

A schematic view of the PANDA detector is shown in Fig. \ref{fig:det}.
The antiprotons circulating in the HESR hit an internal hydrogen target (either pellet
or cluster jet), while for the nuclear part of the experimental program wire
or fiber targets will be used. The apparatus consists of a central detector, called
Target Spectrometer (TS) and a Forward Spectrometer (FS).

The TS, for the measurement of particles emitted at laboratory angles larger
than 5$^\circ$, will be located inside a solenoidal magnet which provides a field
of 2\ T. 
Its main components will be a microvertex silicon detector, a central tracker
(either a straw tube detector or a time projection chamber), an inner time-of-flight
telescope, a cylindrical DIRC (Detector of Internally Reflected Light) for particle
identification, an electromagnetic calorimeter consisting of PbWO$_4$ crystals, a 
set of muon counters and of multiwire drift chambers.

The FS will detect particles emitted at polar angles below 10$^\circ$ in the
horizontal and 5$^\circ$ in the vertical direction. 
It will consist of a 2 T$\cdot$m dipole magnet, with tracking detectors (straw
tubes or multiwire chambers) before and after for charged particle tracking.
Particle identification will be achieved by means of \v{C}erenkov and time-of-flight
detectors. Other components of the FS are and electromagnetic and a hadron 
calorimeter.

All detector components are currently being developed within a very active R\&D
program. This continued development implies that the choice has not yet been
finalized for all detector elements.
\section{Conclusions}

The availability of high-intensity, cooled antiproton beams at FAIR
will make it possible to perform a very rich experimental program.

The PANDA experiment will perform high-precision hadron spectroscopy
from $\sqrt{s} = 2.25\ $GeV to $\sqrt{s} = 5.5\ $GeV and produce 
a wealth of new results:
\begin{itemize}
\item precision measurement of the parameters of all charmonium states,
both below and above open charm threshold, with the possible discovery
of the missing states (e.g. the D-wave states), which will lead to a full
understanding of the charmonium spectrum;
\item the observation/discovery of glueballs and hybrids, particularly
in the mass range between 3 and 5 GeV/c$^2$, yielding new insights into
the structure of the QCD vacuum;
\item the measurement of mass shifts of charmonium and open charm mesons
in nuclear matter, related to the partial restoration of QCD chiral
symmetry in a dense nuclear medium;
\item open charm spectroscopy ($D$ and $D_s$ spectra).
\end{itemize}

All these new measurements
will make it possible to achieve a very significant progress in our
understanding of QCD and the strong interaction.
We are looking forward to many years of exciting hadron physics at FAIR.

\end{document}